УДК 537.534

# ЭВОЛЮЦИЯ ЭНЕРГЕТИЧЕСКИХ И УГЛОВЫХ РАСПРЕДЕЛЕНИЙ ЭМИТИРОВАННЫХ АТОМОВ С ИЗМЕНЕНИЕМ АТОМНОГО НОМЕРА ВЕЩЕСТВА МИШЕНИ

**© 2019 г. В. Н. Самойлов[a,*], А. И. Мусин[b]**

*[a]Московский государственный университет имени М.В. Ломоносова,*

*физический факультет, Москва, Россия*

*[b]Московский государственный областной университет,*

*физико-математический факультет, 105005, Москва, Россия*

**e-mail: samoilov@polly.phys.msu.ru*

С помощью компьютерного моделирования методом молекулярной динамики исследована эмиссия атомов с грани (001) ряда реальных и модельных монокристаллов. Изучена эволюция распределений распыленных с поверхности атомов по энергии с разрешением по полярному и азимутальному углам при изменении атомного номера вещества мишени. При низких энергиях максимум перефокусированных распыленных атомов является более чувствительным к изменению атомного номера вещества мишени, чем максимум фокусированных атомов. Также изучена эволюция полярных угловых распределений распыленных с поверхности атомов с разрешением по энергии и азимутальному углу при изменении атомного номера вещества мишени. Максимумы фокусированных и перефокусированных распыленных атомов являются очень чувствительными к изменению атомного номера вещества мишени. Наблюдаемые сдвиги максимумов связаны с усилением эффекта блокировки с увеличением атомного номера вещества мишени.

# ВВЕДЕНИЕ

В последние годы интерес к эффектам, протекающим при ионной бомбардировке поверхности мишеней, в том числе к распылению твердых тел ионной бомбардировкой значительно возрос. Это связано с большим числом приложений, в которых распыление при ионной бомбардировке играет заметную роль [1]. Отметим роль процесса распыления в современных методах анализа структуры и элементного состава поверхности: вторичной ионной масс-спектрометрии (ВИМС) и масс-спектрометрии вторичных нейтральных частиц (ВНМС), роль процессов, протекающих при ионной бомбардировке, в разрушении первой стенки реакторов. Также этот интерес обусловлен развитием теоретических моделей и, особенно компьютерного моделирования, которые с успехом описывают наблюдающиеся экспериментально особенности и предсказывают новые закономерности процесса распыления.

Прогресс в исследовании процесса распыления тесно связан с развитием как экспериментальных, так и теоретических работ. При этом роль компьютерного моделирования в выяснении механизмов, определяющих особенности, наблюдаемые в экспериментах по ионной бомбардировке поверхности твердых тел, оказывается весьма значительной. Можно отметить недавние успешные исследования сегрегации [2–6], молекулярно-динамические расчеты распыления монокристаллов и поликристаллов разной структуры и состава [7–9], молекулярно-динамические расчеты взаимодействия кластерных ионов с поверхностью и распыления ионами кластеров [10–12], аналитические и компьютерные расчеты самораспыления [13, 14], аналитические расчеты и моделирование распыления двухкомпонентных мишеней [15–21].

Тем не менее, именно в распылении монокристаллов обнаруживаются особенности, не усредненные по углам вылета, как это случается при исследовании распыления поликристаллов. Картина явления с учетом кристаллической структуры и упорядоченности поверхностного атомного слоя монокристалла на стадии эмиссии атомов с поверхности оказывается наиболее сложной. Поэтому анализ механизмов распыления поверхности монокристаллов представляется наиболее актуальным. Одними из основных вопросов остаются механизмы формирования анизотропии двумерного углового распределения атомов, распыленных с поверхности низкоиндексных граней монокристаллов под действием ионной бомбардировки, в том числе, с разрешением по энергии, и механизмы формирования особенностей распределения распыленных атомов по энергии с разрешением по полярному и азимутальному углам вылета.

К настоящему времени опубликовано большое число исследований анизотропии двумерного углового распределения атомов, распыленных с поверхности низкоиндексных граней монокристалла под действием ионной бомбардировки, являющейся одним из сложных эффектов, отражающих анизотропию структуры поверхности кристаллов. Картина углового распределения распыленных атомов чувствительна к типу облучаемой ионами грани кристалла [22–24]. Очень важным в теории распыления является вопрос, какой механизм является ответственным за формирование особенностей двумерного углового распределения атомов, распыленных с поверхности монокристалла, с разрешением по энергии. В [25–31] представлено обсуждение механизмов фокусировки атомов, распыленных с поверхности низкоиндексных граней монокристалла, по полярному и азимутальному углам, в том числе с разрешением по энергии.

Для описания формирования особенностей полярного углового распределения распыленных атомов с разрешением по энергии в [26, 32] была предложена классификация распыленных атомов с разделением на сильно блокированные по полярному углу и остальные распыленные атомы. Сильно блокированные распыленные атомы составляют группу распыленных атомов, для которых отклонение по полярному углу в сторону нормали к поверхности вследствие их рассеяния на атомах поверхности велико. Оно больше отклонения в сторону от нормали к поверхности, которое происходит вследствие преломления траектории распыленного атома на плоском потенциальном барьере.

Отметим также ряд важных для настоящей работы результатов исследований. В работе [33] было обнаружено, что при малых энергиях ионов, бомбардирующих поверхность грани (001) Ni, распределение распыленных атомов по полярному углу в каскаде столкновений, пересекающем поверхность, не имеет максимумов вблизи плотноупакованных направлений <011>. При этом в угловом распределении распыленных атомов наблюдались максимумы эмиссии, которые по своей угловой ширине и направлениям формирования соответствовали наблюдаемым пятнам Венера. В [34] было показано, что блокировка траекторий эмитируемых атомов в сторону меньших полярных углов и по азимутальному углу является одним из основных механизмов формирования наблюдаемых пятен Венера.

В расчетах эмиссии атомов с поверхности граней (001) Ni и (111) Ni, в частности, с разрешением по энергии, наблюдались максимумы в двумерном угловом распределении распыленных атомов, которые по своей угловой ширине и направлениям формирования соответствовали экспериментально наблюдаемым максимумам эмиссии – пятнам Венера [26]. Таким образом, формирование экспериментально наблюдаемых пятен Венера в двумерном угловом распределении атомов, распыленных с поверхности монокристалла, оказалось возможным объяснить действием только поверхностного механизма фокусировки. На стадии эмиссии происходит такое значительное перераспределение вылетающих атомов по углам и энергии, что стадия эмиссии играет важную роль в формировании углового и энергетического распределений распыленных атомов.

Для описания формирования особенностей распределения распыленных атомов по полярному и азимутальному углам с разрешением по энергии в [29–31] предложена классификация распыленных атомов с разделением на “собственные” по азимутальному углу и “несобственные” атомы: фокусированные и перефокусированные. При эмиссии атомов с поверхности грани (001) монокристалла с ГЦК-структурой решетки для несимметричных относительно направления <010> интервалов азимутального угла $\varphi$ формирование сигнала распыленных атомов происходит за счет “собственных” атомов, начальный угол вылета которых $\varphi_0$ принадлежит интервалу углов $\varphi$, и фокусировки “несобственных” атомов. “Несобственные” атомы, в свою очередь, состоят из фокусированных атомов, рассеянных на ближайшем атоме линзы из двух ближайших к эмитируемому атому атомов поверхности, и перефокусированных атомов, рассеянных на дальнем атоме линзы (рис. 1). Для фокусированных атомов угол $\varphi$ и угол $\varphi_0$ лежат по одну сторону от направления <010> на центр линзы из двух ближайших к эмитируемому атому атомов поверхности, для перефокусированных атомов – по разные стороны от этого направления. Таким образом, фокусировка атомов идет к центру линзы из двух атомов, а перефокусировка – через центр линзы из двух атомов поверхности. Эффект перефокусировки был обнаружен в [25, 33] и исследован в ряде работ, например, в [29–31].

Увеличение атомного номера Z вещества мишени приводит к увеличению жесткости потенциала взаимодействия двух атомов и (вследствие этого) к большему отклонению эмитируемых атомов в сторону нормали к поверхности в процессе вылета. Изменения в особенностях эмиссии атомов при переходе от вылета атомов с поверхности грани (001) Ni к вылету с поверхности грани (001) Au исследовались в [28, 29].

В настоящей работе ставилась задача изучить вклад фокусированных и перефокусированных атомов в формирование распределений распыленных атомов по углам и энергии. Целью работы было исследовать изменения распределения распыленных атомов с разрешением по углам и энергии при увеличении атомного номера Z вещества мишени от Z = 28 (Ni) до Z = 79 (Au).

# МОДЕЛЬ РАСЧЕТА

Расчеты были проведены для эмиссии атомов с поверхности грани (001) ряда реальных и модельных монокристаллов с атомными номерами от Z = 28 (Ni) до Z = 79 (Au). Поверхность кристалла моделировалась 20 атомами поверхности, ближайшими к узлу решетки, из которого происходила эмиссия атома (модель 21 атома). Рассеяние атома при вылете с поверхности и классификация эмитированных атомов по азимутальному углу φ представлены на рис. 1. Представлена только линза, состоящая из двух атомов, ближайших к вылетающему атому соседей в плоскости поверхности. Подобная модель использовалась в наших работах [20, 30, 31].

Для расчета эмиссии атомов использовался метод молекулярной динамики. Взаимодействие эмитируемого атома с атомами поверхности в модели описывалось потенциалом отталкивания, а на достаточно большом удалении атома от поверхности был введен плоский потенциальный барьер. В качестве потенциала взаимодействия атом–атом был использован потенциал Борна–Майера:

$$U(r) = A\exp(-r/b) \tag{1}$$

с параметрами $A$ = 23853.96 эВ и $b$ = 0.196 Å для взаимодействия двух атомов Ni из работы [35]. Энергия связи составляла 4.435 эВ.

Таким образом, реальные силы взаимодействия атомов (отталкивания–притяжения) заменялись силами отталкивания, а притяжение к кристаллу моделировалось плоским потенциальным барьером при большом удалении атома от поверхности. Использование потенциала отталкивания вместо потенциала отталкивания–притяжения дает возможность разделить вклады рассеяния эмитируемого атома на ближайших атомах поверхности и его дальнейшего притяжения к поверхности – преломления на плоском потенциальном барьере – в характеристики распыления.

При переходе к элементам с большим атомным номером Z, например, при переходе от Ni к Au, возрастает константа $A$ в выражении (1) для потенциала взаимодействия атом–атом [35] и увеличивается сечение взаимодействия сталкивающихся атомов. В частности, используя выражения для констант потенциала взаимодействия атом–атом [35], получаем:

$$\frac{A_{Au}}{A_{Ni}} = \left(\frac{Z_{Au}}{Z_{Ni}}\right)^{5/3}. \qquad (2)$$

Согласно (2) при переходе от Ni к Au константа $A$ возрастает в ≈ 5.63 раза. Параметры кристаллической решетки не менялись при переходе к расчетам эмиссии атомов с грани (001) монокристаллов с большим атомным номером Z. Это дало возможность исследовать изменения особенностей эмиссии при изменении только одного параметра – атомного номера вещества мишени Z. При переходе к элементам с большим атомным номером Z использовалась одинаковая величина энергии связи атомов на поверхности, равная энергии сублимации для никеля 4.435 эВ. Этим мы исключаем влияние разницы энергий связи атомов для мишеней с большим атомным номером Z на особенности эмиссии атомов с поверхности.

Атом выбивался из узла на поверхности с энергией $E_0$ под углами $\vartheta_0$ (начальный полярный угол, отсчитывался от нормали к поверхности) и $\varphi_0$ (начальный азимутальный угол, $\varphi_0 = 90°$ соответствовал направлению <010> на центр линзы из двух ближайших к эмитируемому атому атомов поверхности). Начальная энергия $E_0$ изменялась от 0.5 эВ до 100 эВ. Шаг по $E_0$ составлял 0.01 эВ. Шаг по $\varphi_0$ был равен 0.5°, шаг по $1 - \cos\vartheta_0$ составлял 1/450. Было использовано начальное распределение эмитируемых атомов по углам и энергии $\cos\vartheta_0/E_0^2$ [36, 37]. Таким образом, распределение эмитируемых атомов по начальному азимутальному углу $\varphi_0$ было изотропным.

Считалось, что распыление происходит только за счет атомов поверхностного слоя. Это допущение вполне оправдано ввиду того, что для мишеней, состоящих из средних по массе и тяжелых атомов, вклад атомов поверхностного слоя в распыление является доминирующим (88.6% для случая ионной бомбардировки Cu [38], 82% для случая ионной бомбардировки Mo [39]). Обсуждение некоторых особенностей и корректности модели, используемой в настоящей работе, приведено также в работе [25].

При выполнении расчетов исследован вопрос, каким образом взаимодействие эмитируемых атомов с атомами поверхности кристалла в процессе вылета влияет на особенности наблюдаемых распределений распыленных атомов по углам и энергии.

Расчеты были выполнены с использованием оборудования Центра коллективного пользования сверхвысокопроизводительными вычислительными ресурсами МГУ имени М.В. Ломоносова [40, 41].

# РЕЗУЛЬТАТЫ И ИХ ОБСУЖДЕНИЕ

*Эволюция распределений распыленных атомов по энергии с изменением атомного номера вещества мишени*

Изучено изменение распределений распыленных атомов по энергии с одновременным разрешением по полярному $\vartheta$ и азимутальному $\varphi$ углам при изменении атомного номера Z вещества мишени. Для несимметричных относительно направления <010> интервалов угла $\varphi$ в распределениях распыленных атомов по энергии различаются максимумы для фокусированных и перефокусированных атомов [30, 31]. При сравнительно низких энергиях (ниже ~8 эВ) в распределении распыленных атомов по энергии при углах наблюдения $\varphi$ [76.5°, 79.5°] и $\vartheta$ [49.9°, 51.5°] для Z = 28 (Ni) наблюдаются максимум фокусированных атомов при энергии ~3 эВ и максимум перефокусированных атомов при энергии ~6 эВ (рис. 2).

При увеличении Z до 29 (Cu) оба максимума смещаются менее чем на 0.5 эВ в сторону меньших энергий. При увеличении Z до 47 (Ag) и далее до 79 (Au) обнаружен сдвиг максимума фокусированных атомов на 2.5 эВ в сторону меньших энергий, в то время как максимум перефокусированных атомов смещается более чем на 5 эВ в ту же сторону. Таким образом, при низких энергиях максимум перефокусированных распыленных атомов является более чувствительным к изменению атомного номера вещества мишени, чем максимум фокусированных атомов.

При сравнительно высоких энергиях (выше ~40 эВ) в распределении фокусированных распыленных атомов по энергии при углах наблюдения $\varphi$ [76.5°, 79.5°] и $\vartheta$ [49.9°, 51.5°] для Z = 28 (Ni) наблюдается максимум при энергии ~64 эВ (рис. 3). Это высокоэнергетический максимум в распределении распыленных атомов по энергии с разрешением по азимутальному и полярному углам. Высокоэнергетический пик сравнительно мал по величине, однако он является достаточно протяженным и выраженным и представляет собой новую важную особенность распыления монокристаллов с угловым разрешением.

Этот максимум был обнаружен и исследован в расчетах распределений распыленных атомов по энергии с угловым разрешением [42]. Было обнаружено, что этот пик практически целиком образован распыленными атомами, сильно блокированными в сторону нормали к поверхности на стадии эмиссии, распыленными под углами $\vartheta < \vartheta_0$ (такой вклад отсутствует в моделях распыления без учета атомной дискретности поверхности на стадии эмиссии). Было показано, что эти атомы эмитировались с поверхности под большими полярными углами $\vartheta_0$ от нормали к поверхности.

Для этого азимутального направления φ [76.5°, 79.5°] наблюдается важная особенность, которая уже наблюдалась в работе [42], состоящая в том, что в некоторых направлениях не распыляются атомы с энергиями между низкоэнергетическим и высокоэнергетическим максимумами распределения распыленных атомов по энергии (рис. 2, 3). Так, для выбранных интервалов азимутального и полярного углов полностью отсутствуют атомы, распыленные с энергиями в диапазоне от 7.5 до 45 эВ. В [42], эти энергии были названы “запрещенными”. Атомы в этом диапазоне энергий сильно отклоняются в сторону нормали к поверхности в процессе вылета и наблюдаются при меньших полярных углах ϑ. Запрещенные” энергии соответствуют области тени, то есть полному отсутствию распыленных атомов при определенных энергии и полярных углах, возникающей вследствие рассеяния эмитируемых атомов на атомах поверхности в сторону нормали к поверхности в процессе вылета.

При увеличении Z до 29 (Cu) высокоэнергетический максимум смещается на ~6 эВ до ~70 эВ, то есть сдвиг максимума происходит в сторону больших энергий. Таким образом, при высоких энергиях максимум фокусированных распыленных атомов оказывается чувствительным даже к небольшому изменению атомного номера вещества мишени. Этот сдвиг происходит в другую сторону по отношению к сдвигу низкоэнергетического максимума. В настоящей работе было показано, что сдвиг высокоэнергетического максимума в сторону больших энергий с увеличением атомного номера вещества мишени Z определяется рассеянием эмитированного атома на ближайших атомах поверхности и связан с увеличением угловых размеров тени за рассеивающими атомами поверхности с увеличением атомного номера вещества мишени.

При дальнейшем увеличении Z до 47 (Ag) и далее до 79 (Au) оказалось, что в этих интервалах азимутального и полярного углов наблюдения высокоэнергетического максимума распыленных атомов не наблюдается. Это связано с сильной блокировкой распыленных атомов в сторону нормали к поверхности, в связи с этим сигнал распыленных атомов при этих углах наблюдения и энергиях становится равным нулю.

*Изменение фокусировки распыленных атомов с увеличением атомного номера вещества мишени*

Была изучена эволюция полярных угловых распределений распыленных атомов с одновременным разрешением по энергии *E* и азимутальному углу φ при изменении атомного номера Z вещества мишени. Для несимметричных относительно направления <010> интервалов угла φ в распределениях распыленных атомов по полярному углу ϑ различаются максимумы для фокусированных и перефокусированных атомов [30, 31]. В полярном угловом распределении распыленных атомов для энергии $E$ = 2.0 ± 0.4 эВ и углов φ [76.5°, 79.5°] для Z = 28 (Ni) наблюдаются максимум фокусированных атомов при ϑ ~55.5° и максимум перефокусированных атомов при ϑ ~72.5° (рис. 4).

Распределение распыленных атомов по полярному углу для фиксированных интервалов энергии и азимутального угла состоит из вкладов фокусированных атомов (левый максимум), перефокусированных атомов (правый максимум) (рис. 4б) и вклада “собственных” атомов вблизи нормали к поверхности. Распределение распыленных атомов по полярному углу для того же интервала азимутальных углов без разрешения по энергии имеет один максимум, образованный фокусированными и перефокусированными атомами.

При увеличении Z до 29 (Cu) оба максимума незначительно смещаются в сторону нормали к поверхности. При увеличении Z до 47 (Ag) и далее до 79 (Au) обнаружен сдвиг максимума фокусированных атомов в сторону нормали до ϑ ~14.8°, максимум перефокусированных атомов смещается в ту же сторону до ϑ ~33.6°. Таким образом, максимумы фокусированных и перефокусированных распыленных атомов в полярном угловом распределении распыленных атомов при низкой энергии являются очень чувствительными к изменению атомного номера вещества мишени. Эти сдвиги связаны с усилением эффекта блокировки с увеличением Z.

*Эволюция распределений распыленных атомов с разрешением по энергии и полярному углу с увеличением атомного номера вещества мишени*

В распределениях распыленных атомов с одновременным разрешением по энергии и полярному углу для фиксированных интервалов углов φ отчетливо различаются отдельные хребты – максимумы распределений для фокусированных и перефокусированных атомов (рис. 5). Верхний хребет образован в основном фокусированными атомами, нижний – только перефокусированными атомами. Максимум распределения перефокусированных атомов наблюдается в области энергии и полярных углов, при которых нет вылета других групп атомов. Перефокусированные атомы на 100% формируют наблюдаемый сигнал. Таким образом, в экспериментах с разрешением по углам и энергии оказывается принципиально возможным выделить отдельно сигнал только перефокусированных атомов.

Если в распределении распыленных атомов с одновременным разрешением по энергии и полярному углу для Z = 29 (Cu), представленном на рис. 5а, мысленно выделить по горизонтали интервал углов ϑ [49.9°, 51.5°], то становится понятной структура наблюдаемых максимумов на рис. 2а и 3. При этом видно, что между низкоэнергетической частью распре-деления по энергии и высокоэнергетической частью наблюдается область тени, то есть полное отсутствие распыленных атомов при энергии $E$ = 7.3–49 эВ.

При переходе от Z = 28 (Ni) к Z = 29 (Cu) вследствие увеличения сечения взаимодействия атомов с увеличением атомного номера Z вещества мишени происходит сдвиг всей структуры распределения, представленного на рис. 5а, в сторону нормали к поверхности (вверх) и в сторону меньших энергий распыленных атомов (влево). Становится понятным, почему при этом происходит сдвиг низкоэнергетических максимумов фокусированных и перефокусированных атомов в сторону меньших энергий (рис. 2а), сдвиг высокоэнергетического максимума в сторону больших энергий (рис. 3) и расширение области “запрещенных” энергий между низкоэнергетической частью и высокоэнергетической частью распределения по энергии, то есть увеличение области тени.

Если в распределении распыленных атомов с одновременным разрешением по энергии и полярному углу для Z = 29 (Cu), представленном на рис. 5а, мысленно выделить по вертикали интервал энергии $E = 1.2 \pm 0.4$ эВ и дальше увеличивать энергию, то в распределении по $1 - \cos\vartheta$ для фокусированных распыленных атомов будет наблюдаться немонотонный сдвиг максимума распределения. Сначала с увеличением энергии до $E = 10.0 \pm 0.4$ эВ максимум распределения смещается в сторону нормали к поверхности, а при дальнейшем увеличении энергии – в сторону от нормали к поверхности.

Этот немонотонный сдвиг максимума полярного углового распределения распыленных атомов обусловлен следующими причинами. При формировании положения максимума полярного углового распределения распыленных атомов с изменением энергии $E$ происходит конкуренция двух факторов: блокировки атомов в сторону нормали к поверхности и преломления на плоском потенциальном барьере. С ростом энергии действие эффекта блокировки, то есть отклонение эмитируемого атома в сторону нормали к поверхности, уменьшается вследствие уменьшения сечения взаимодействия атом–атом, что приводит к сокращению угловых размеров тени от линзы из двух рассеивающих центров [25, 26]. С другой стороны, с ростом энергии плоский барьер все менее эффективно отклоняет атомы в сторону от нормали к поверхности. В настоящем случае при сравнительно малых (менее 10 эВ) энергиях $E$ (и, соответственно, малых энергиях $E_0$) энергия атомов перед плоским потенциальным барьером оказывается соизмеримой с величиной энергии связи, что приводит к сильному влиянию плоского барьера. В результате в этом диапазоне энергий $E$ уменьшение энергии приводит к более сильному преломлению атомов на потенциальном барьере и, следовательно, к сдвигу максимума полярного углового распределения в сторону от нормали к поверхности.

В диапазоне энергий $E > 10$ эВ с увеличением энергии происходит сдвиг максимума полярного углового распределения в сторону от нормали к поверхности. Этот сдвиг более слабый и связан с сокращением угловых размеров тени от линзы из двух рассеивающих атомов в плоскости поверхности [25, 26]. Немонотонный сдвиг максимума полярного углового распределения распыленных атомов с увеличением энергии наблюдался также в расчетах с применением полной модели молекулярной динамики (с учетом взаимодействия бомбардирующих ионов с поверхностью) [32].

При переходе от Z = 29 (Cu) к Z = 47 (Ag) вследствие увеличения сечения взаимодействия атомов с увеличением атомного номера Z вещества мишени происходит довольно резкий сдвиг всей структуры распределения, представленного на рис. 5а, в сторону нормали к поверхности (вверх) и в сторону меньших энергий распыленных атомов (влево), рис. 5б. Становится понятным, почему при этом происходит довольно резкий сдвиг низкоэнергетических максимумов фокусированных и перефокусированных атомов в сторону меньших энергий (рис. 2). При этом высокоэнергетического максимума при этих углах наблюдения $\varphi$ [76.5°, 79.5°] и $\vartheta$ [49.9°, 51.5°] для Z = 47 (Ag) уже не наблюдается (рис. 3) из-за расширения области “запрещенных” энергий и резкого увеличения области тени за рассеивающими атомами поверхности. Кроме того, ясно, почему при энергии $E = 2.0 \pm 0.4$ эВ при переходе к Z = 47 (Ag) наблюдается сильный сдвиг максимумов фокусированных и перефокусированных атомов в сторону нормали к поверхности (рис. 4). Отметим, что и для высоких энергий $E > 30$ эВ сдвиг максимума полярного углового распределения распыленных атомов с увеличением атомного номера Z вещества мишени происходит также в сторону нормали к поверхности (рис. 5). Это связано с возрастанием сечения взаимодействия атомов с увеличением атомного номера Z вещества мишени.

## ЗАКЛЮЧЕНИЕ

С помощью модели молекулярной динамики исследованы особенности эмиссии атомов с грани (001) ряда реальных и модельных монокристаллов. Обсуждаются механизмы формирования распределений распыленных атомов с разрешением одновременно по энергии, азимутальному и полярному углам.

Исследована эволюция распределений распыленных атомов по энергии с одновременным разрешением по полярному и азимутальному углам с изменением атомного номера вещества мишени. Показано, что в распределении распыленных атомов по энергии с увеличением атомного номера вещества мишени происходит заметный сдвиг низкоэнергетических максимумов фокусированных и перефокусированных атомов в сторону меньших энергий. При этом максимум перефокусированных распыленных атомов является более чувствительным к изменению атомного номера вещества мишени, чем максимум фокусированных атомов. Обнаружено, что высокоэнергетический максимум, сформированный фокусированными сильно блокированными атомами, смещается в сторону больших энергий с увеличением атомного номера вещества мишени.

Выявлено изменение распределения распыленных атомов по полярному углу с одновременным разрешением по энергии и азимутальному углу при изменении атомного номера вещества мишени. При низких энергиях максимумы фокусированных и перефокусированных распыленных атомов являются очень чувствительными к изменению атомного номера вещества мишени и смещаются с увеличением атомного номера в сторону нормали к поверхности. При высоких энергиях максимум распределения распыленных атомов смещается также в сторону нормали к поверхности с увеличением атомного номера вещества мишени. Эти сдвиги связаны с усилением эффекта блокировки с увеличением атомного номера вещества мишени.

Исследовано изменение распределений распыленных атомов с одновременным разрешением по энергии и полярному углу с увеличением атомного номера вещества мишени. Показано, что при переходе к большим атомным номерам Z вещества мишени вследствие увеличения сечения взаимодействия атомов с увеличением атомного номера Z происходит довольно резкий сдвиг всей структуры распределения в сторону нормали к поверхности и в сторону меньших энергий распыленных атомов.

Это становится причиной относительно резкого сдвига низкоэнергетических максимумов фокусированных и перефокусированных атомов в сторону меньших энергий. При этом происходит сдвиг высокоэнергетического максимума, образованного фокусированными сильно блокированными атомами в сторону больших энергий при небольшом изменении Z. Высокоэнергетического максимума при этих углах наблюдения для Z = 47 (Ag) и выше уже не наблюдается из-за расширения области “запрещенных” энергий и резкого увеличения области тени за рассеивающими атомами поверхности.

Возрастание сечения взаимодействия атомов с увеличением атомного номера Z вещества мишени является также причиной сильного сдвига максимумов фокусированных и перефокусированных атомов в распределении распыленных атомов по полярному углу в сторону нормали к поверхности при низкой энергии. Это же является причиной заметного сдвига максимума полярного углового распределения с увеличением атомного номера Z также в сторону нормали к поверхности при высоких энергиях.

# СПИСОК ЛИТЕРАТУРЫ

# EVOLUTION OF ENERGY AND ANGULAR DISTRIBUTIONS OF SPUTTERED ATOMS WITH VARIATION OF ATOMIC NUMBER OF SINGLE CRYSTAL TARGET

**V. N. Samoilov[1, #], A. I. Musin[2]**

[1]*M.V. Lomonosov Moscow State University, Faculty of Physics, Moscow, Russia*

[2]*Moscow Region State University, Faculty of Physics and Mathematics, Moscow, Russia*

[#]*e-mail: samoilov@polly.phys.msu.ru*

Ejection of atoms from (001) surfaces of a number of real and model single crystals is studied by molecular dynamics computer simulation. Evolution of energy distributions of sputtered atoms with polar and azimuthal angle resolution with increase of atomic number of the target is investigated. For low energies the maximum of overfocused atoms is more sensitive to the increase of atomic number of the target than the maximum of focused atoms. Evolution of the polar angular distributions of sputtered atoms with energy and azimuthal angle resolution with increase of atomic number of the target is also studied. Maxima of focused and overfocused atoms are very sensitive to the increase of atomic number of the target. The observed maxima shifts are due to stronger blocking effect with increase of atomic number of the target.

# ПОДПИСИ К РИСУНКАМ

Рис. 1. Рассеяние атома при вылете с поверхности и классификация эмитированных атомов по азимутальному углу φ. Представлена только линза, состоящая из двух атомов, ближайших к вылетающему атому соседей в плоскости поверхности.

Рис. 2. Распределения всех распыленных атомов (а) и только фокусированных распыленных атомов (б) по энергии *E* при углах наблюдения φ [76.5°, 79.5°] и ϑ [49.9°, 51.5°] для атомных номеров Z вещества мишени: 28 (*1*), 29 (*2*), 47 (*3*), 74 (*4*), 79 (*5*).

Рис. 3. Распределения фокусированных распыленных атомов по энергии *E* при углах наблюдения φ [76.5°, 79.5°] и ϑ [49.9°, 51.5°] для атомных номеров Z вещества мишени: 28 (*1*), 29 (*2*). Отчетливо виден сильный сдвиг высокоэнергетического максимума распределения даже при небольшом изменении Z.

Рис. 4. Распределения всех распыленных атомов (а) и только перефокусированных распыленных атомов (б) по 1 – cosϑ при энергии *E* = 2.0 ± 0.4 эВ и углах φ [76.5°, 79.5°] для атомных номеров Z вещества мишени: 28 (*1*), 29 (*2*), 47 (*3*), 74 (*4*), 79 (*5*).

Рис. 5. Распределения распыленных атомов при эмиссии с грани (001) Cu, Z = 29 (а) и (001) Ag, Z = 47 (б) одновременно по 1 – cosϑ и энергии *E* для интервала азимутальных углов φ [76.5°, 79.5°]. Верхний хребет образован в основном фокусированными атомами, нижний – только перефокусированными атомами.

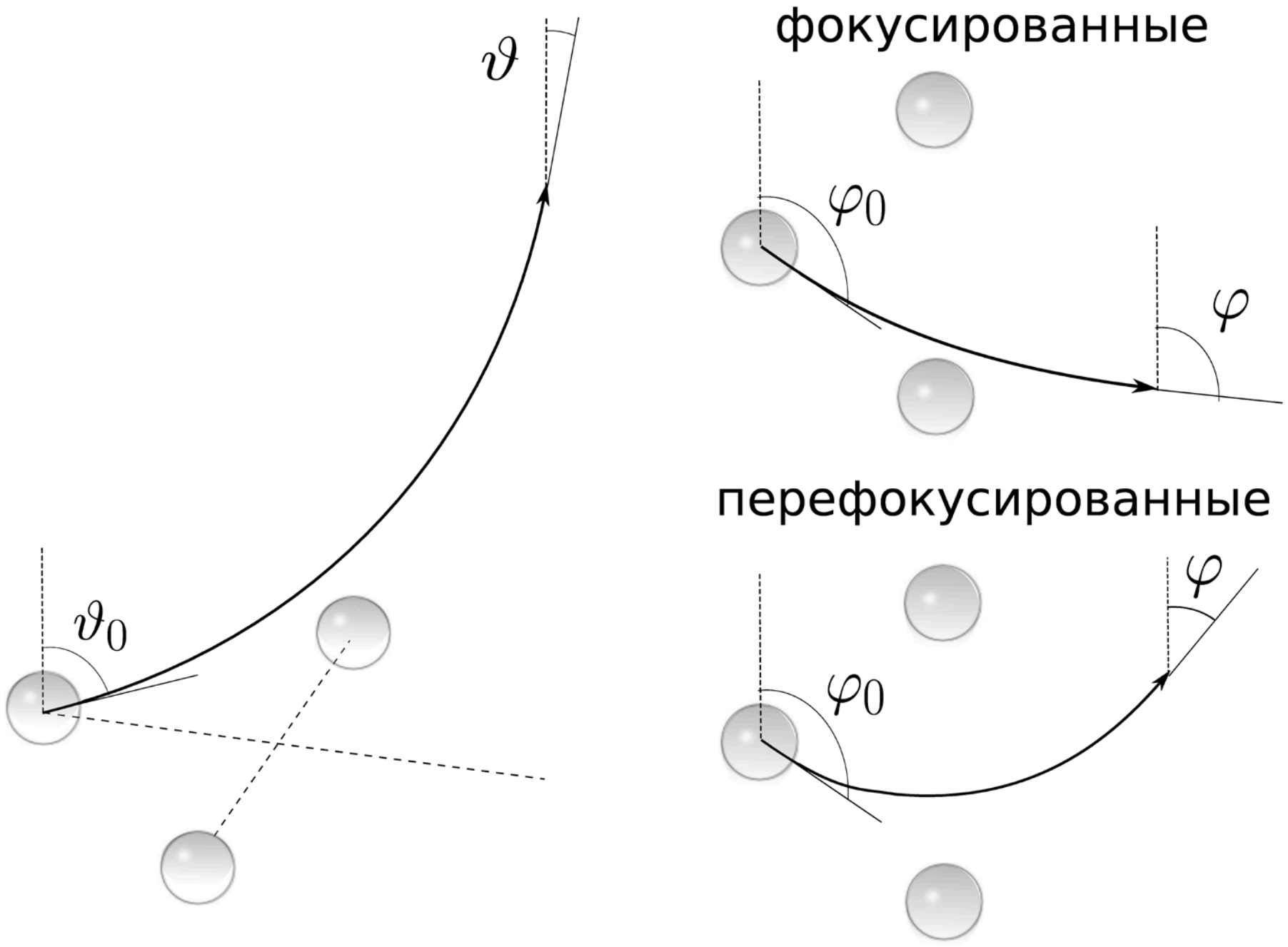

Рис. 1. В. Н. Самойлов, А. И. Мусин _Поверхность

Рис. 1. Рассеяние атома при вылете с поверхности и классификация эмитированных атомов по азимутальному углу φ. Представлена только линза, состоящая из двух атомов, ближайших к вылетающему атому соседей в плоскости поверхности.

(a)

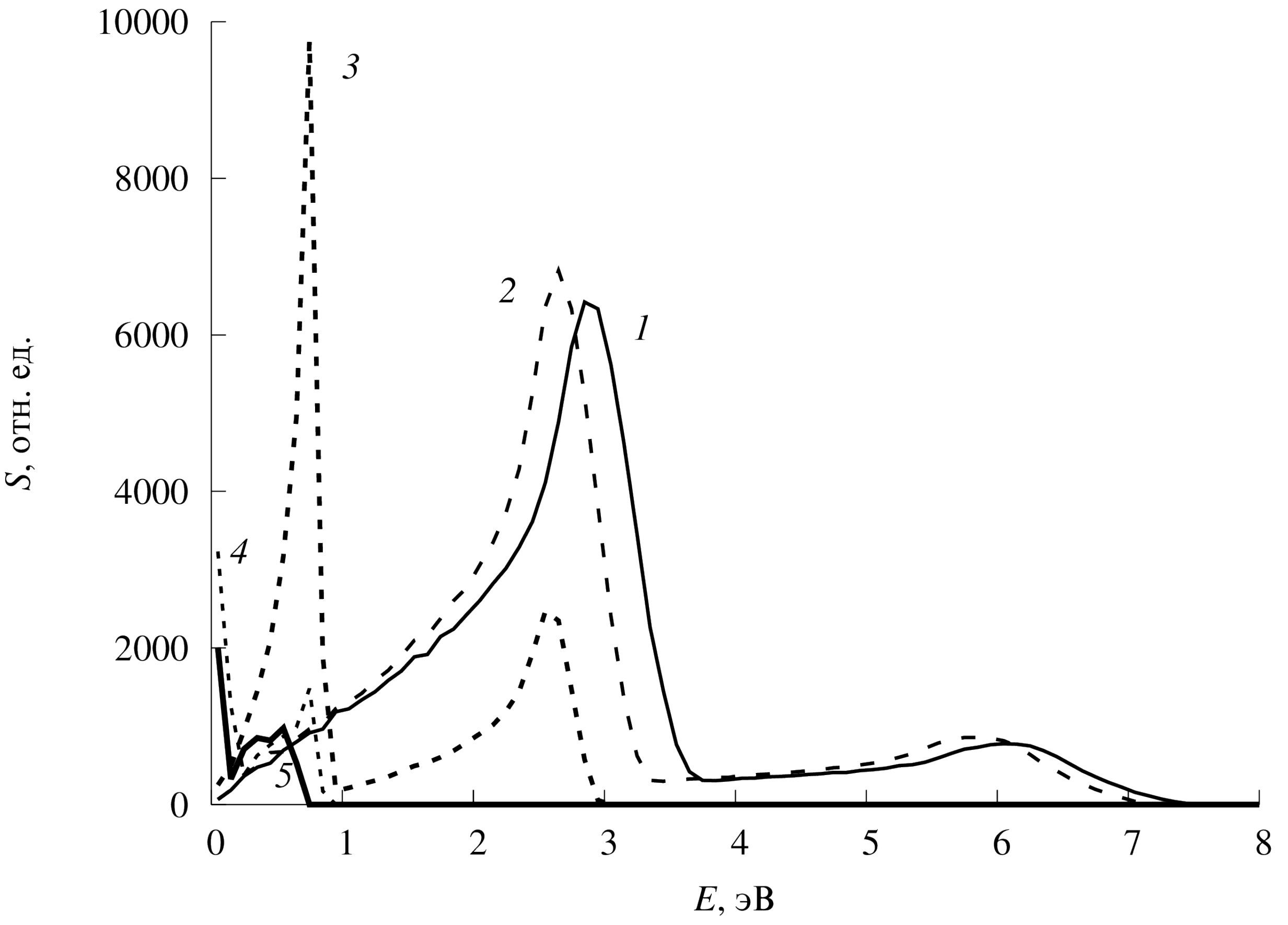

Рис. 2. В. Н. Самойлов, А. И. Мусин _Поверхность

(б)

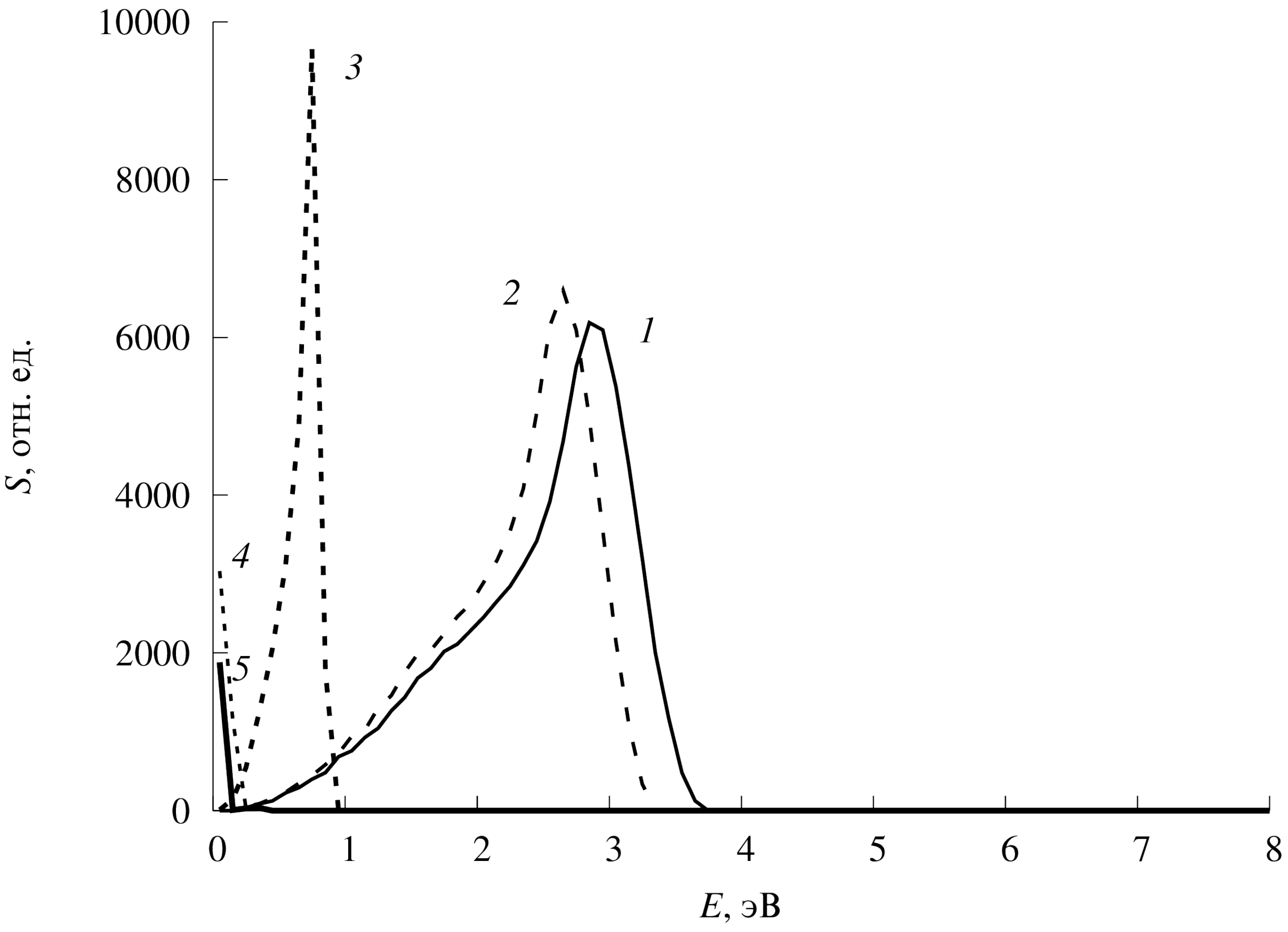

Рис. 2. В. Н. Самойлов, А. И. Мусин _Поверхность

Рис. 2. Распределения всех распыленных атомов (а) и только фокусированных распыленных атомов (б) по энергии *E* при углах наблюдения φ [76.5°, 79.5°] и ϑ [49.9°, 51.5°] для атомных номеров Z вещества мишени: 28 (*1*), 29 (*2*), 47 (*3*), 74 (*4*), 79 (*5*).

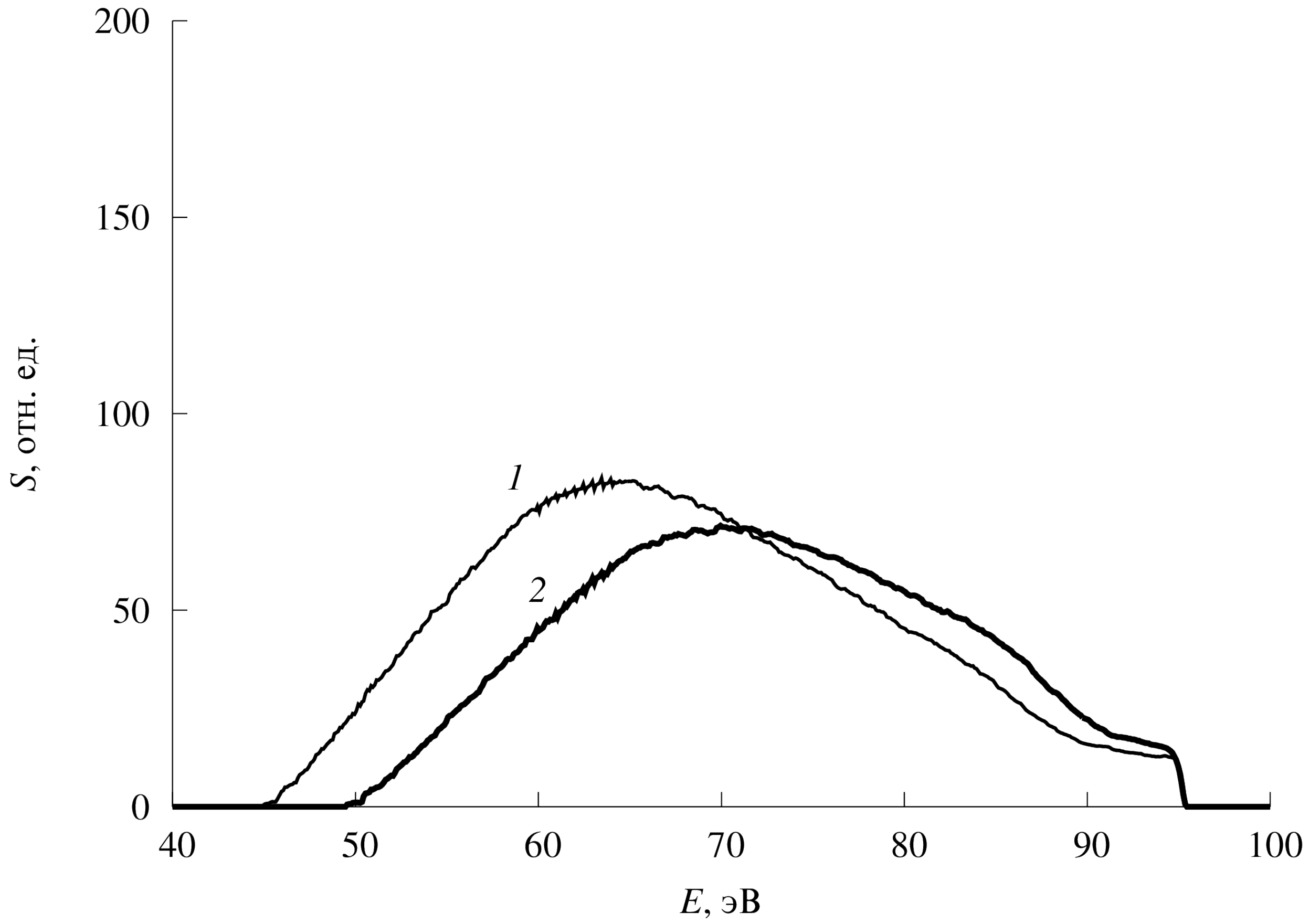

Рис. 3. В. Н. Самойлов, А. И. Мусин _Поверхность

Рис. 3. Распределения фокусированных распыленных атомов по энергии *E* при углах наблюдения φ [76.5°, 79.5°] и ϑ [49.9°, 51.5°] для атомных номеров Z вещества мишени: 28 (*1*), 29 (*2*). Отчетливо виден сильный сдвиг высокоэнергетического максимума распределения даже при небольшом изменении Z.

(a)

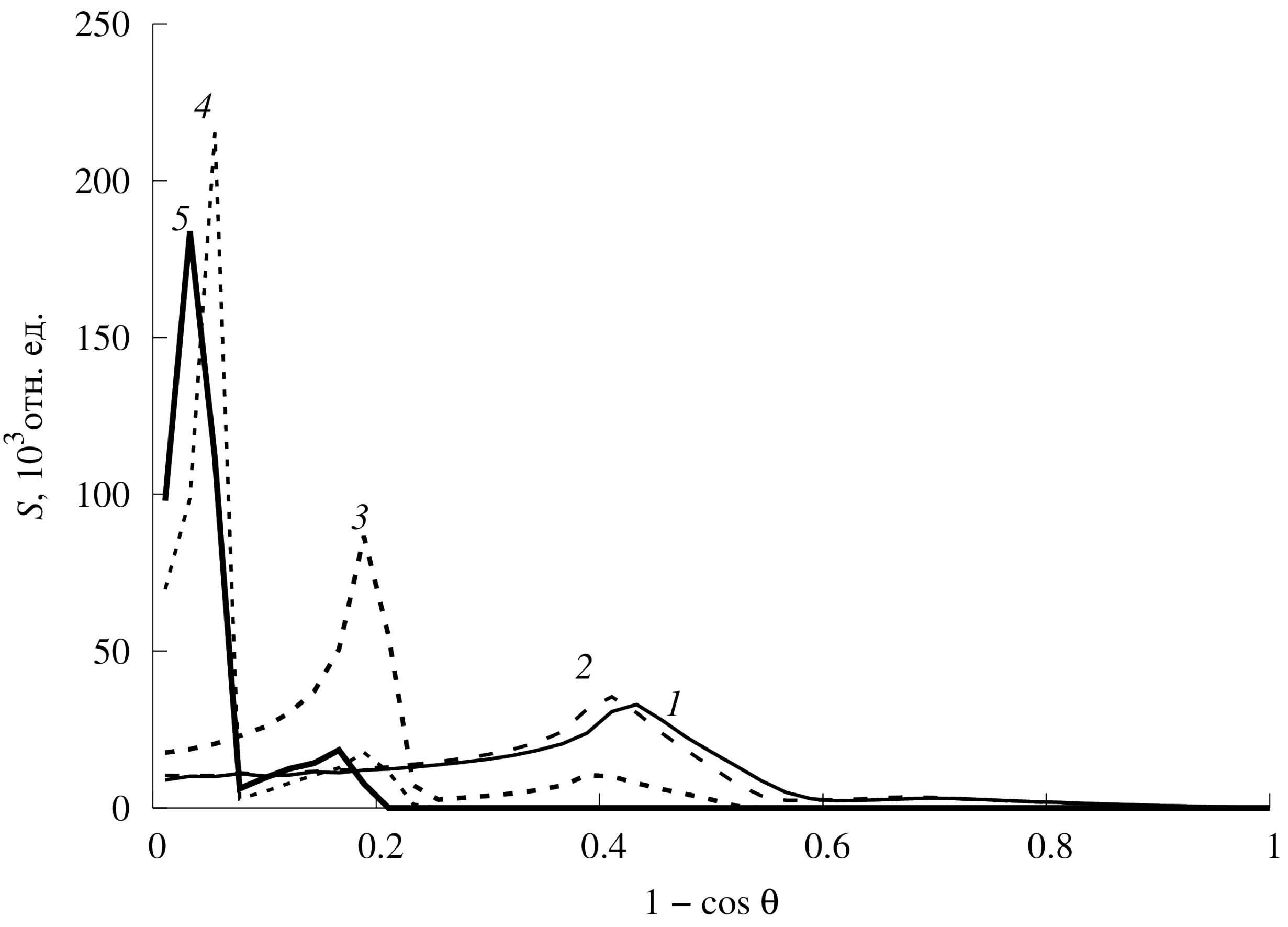

Рис. 4. В. Н. Самойлов, А. И. Мусин _Поверхность

(б)

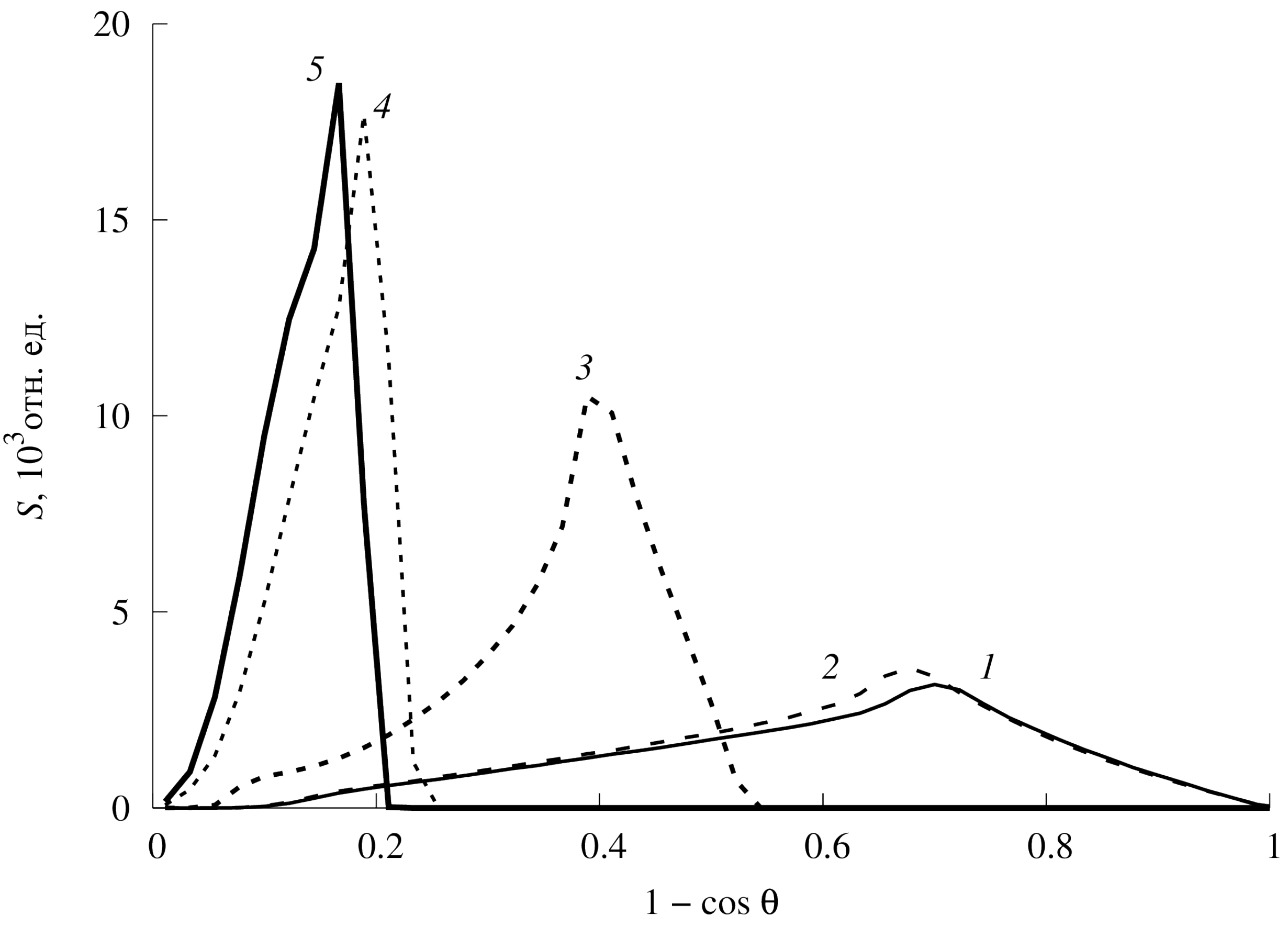

Рис. 4. В. Н. Самойлов, А. И. Мусин _Поверхность

Рис. 4. Распределения всех распыленных атомов (а) и только перефокусированных распыленных атомов (б) по 1 – cosϑ при энергии *E* = 2.0 ± 0.4 эВ и углах φ [76.5°, 79.5°] для атомных номеров Z вещества мишени: 28 (*1*), 29 (*2*), 47 (*3*), 74 (*4*), 79 (*5*).

(a)

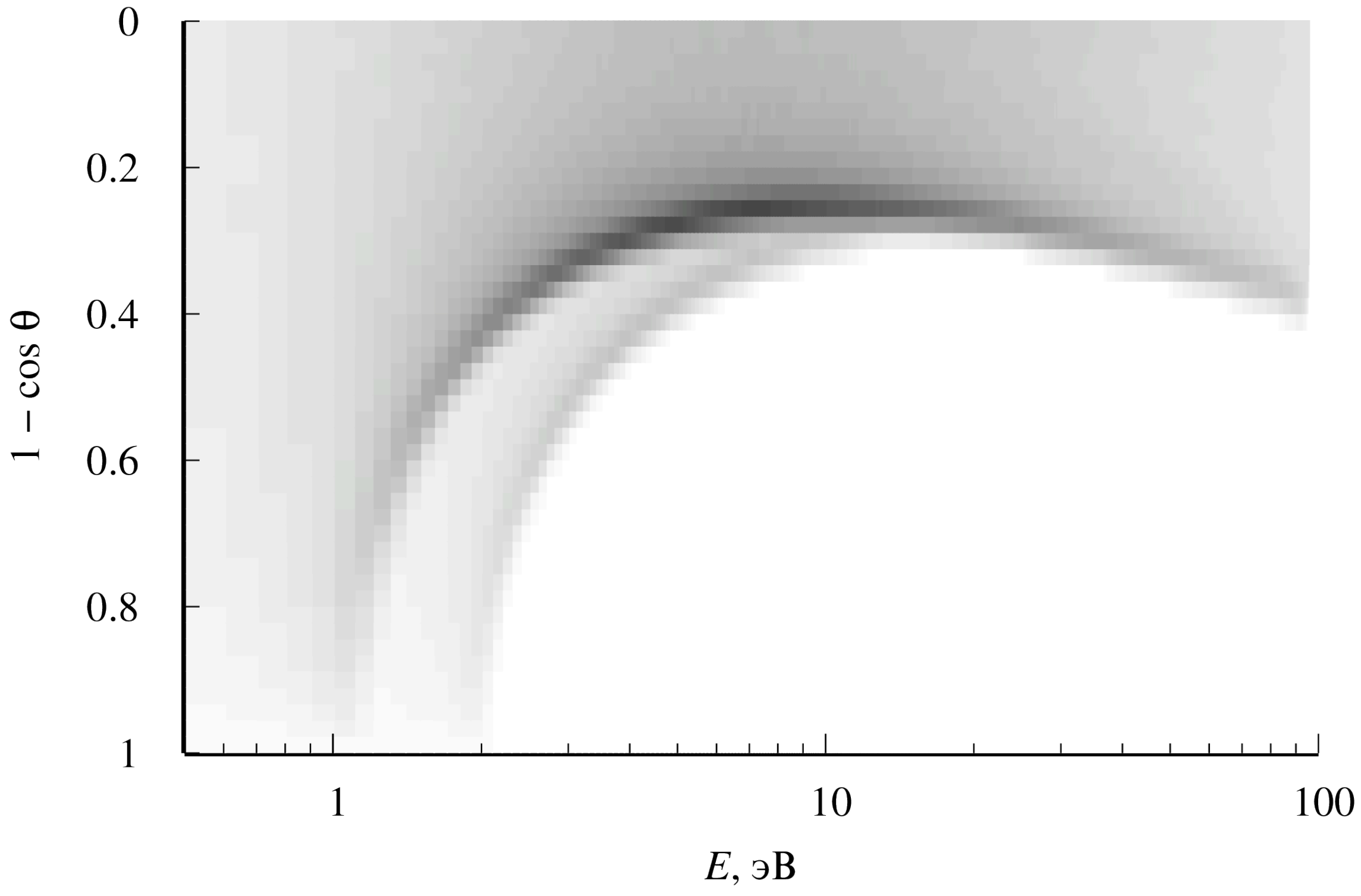

Рис. 5. В. Н. Самойлов, А. И. Мусин _Поверхность

(б)

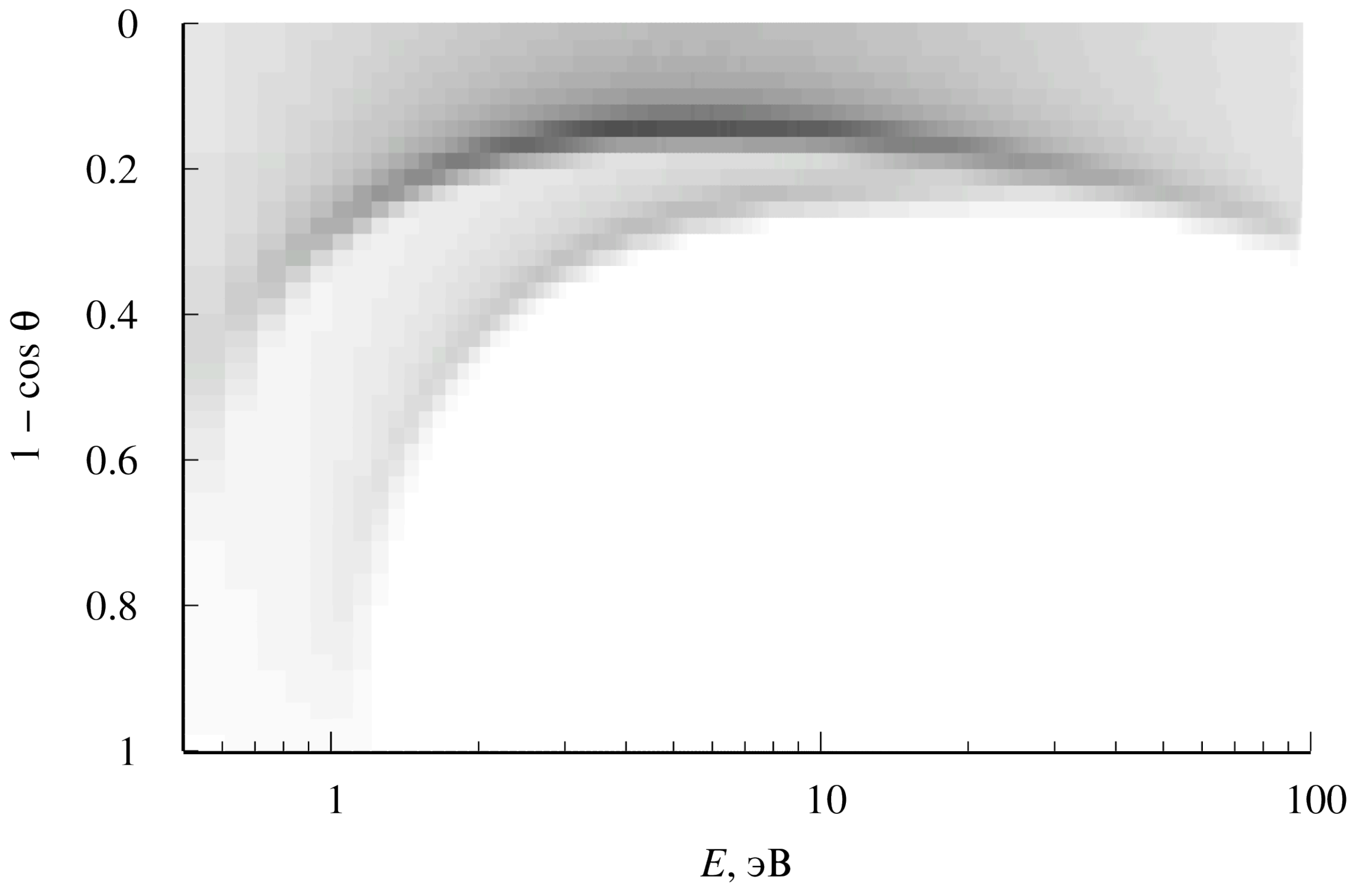

Рис. 5. В. Н. Самойлов, А. И. Мусин _Поверхность

Рис. 5. Распределения распыленных атомов при эмиссии с грани (001) Cu, Z = 29 (а) и (001) Ag, Z = 47 (б) одновременно по 1 – cosϑ и энергии *E* для интервала азимутальных углов φ [76.5°, 79.5°]. Верхний хребет образован в основном фокусированными атомами, нижний – только перефокусированными атомами.